\documentclass{article}
\usepackage{epsfig}

\date{\today}

\usepackage{amsmath}

\newcommand{\ee}{\end{equation}}
\newcommand{\eea}{\end{eqnarray}}
\newcommand{\be}{\begin{equation}}
\newcommand{\bea}{\begin{eqnarray}}

\begin{document}

\title{Deformed vortices in (4+1)-dimensional Einstein-Yang-Mills
theory}
\author{{\large Yves Brihaye \footnote{yves.brihaye@umh.ac.be}}\\
\small{
Facult\'e des Sciences, Universit\'e de Mons-Hainaut,
B-7000 Mons, Belgium }\\
{ }\\
{\large Betti Hartmann \footnote{b.hartmann@iu-bremen.de}}\\
\small{
School of Engineering and Science, International University Bremen,
28725 Bremen, Germany
 }\\
{ }\\
{\large Eugen Radu\footnote{radu@heisenberg1.thphys.may.ie}}\\
\small{
Department of  Mathematical Physics,
National University of Ireland Maynooth, Ireland}}

\date{\today}

\maketitle
\begin{abstract}
We study vortex-type solutions in a (4+1)-dimensional
Einstein-Yang-Mills-SU(2) model.
Assuming all fields to be independent on the extra coordinate,
these solutions correspond in a  four dimensional picture to axially
symmetric multimonopoles, respectively monopole-antimonopole
solutions.
By boosting the five dimensional purely magnetic solutions we find
new configurations
which in four dimensions represent rotating regular nonabelian
solutions with an additional electric charge.
\end{abstract}

\section{Introduction}
Ever since the pioneering work of Kaluza and Klein in the 1920s
\cite{kk},
extra dimensions
are intensively discussed in physics. Specifically the result that
string theory is only consistent in 10, respectively 26 dimensions
for superstring and bosonic string theory has boosted the interest.
In string theory \cite{pol}, the extra dimensions are usually
compactified at
the Planck length, while in recent years so-called brane world
models have emerged which can have non-compact and even infinite
extra dimensions
\cite{brane}.

When discussing solutions in higher dimensional theories, two
approaches
seem possible: either to study solutions with a specific symmetry in
the
full dimensions or to study solutions with a symmetry in four
dimensions
which are then trivially extended into the extra dimensions.

Recently, gravitating solutions including non-abelian gauge fields
have been
discussed in this context.
In \cite{Brihaye:2002jg}, (4+1)-dimensional generalizations
of the Bartnik-McKinnon solutions \cite{Bartnik:1988am}
with an SO(4) symmetry group, have been
studied in Einstein-Yang-Mills (EYM) theory.
However, as was proven in \cite{Volkov:2001tb} and demonstrated
numerically
in \cite{Brihaye:2002jg}, in this case there are
no finite energy solutions unless one considers
the inclusion of higher order curvature and/or Born-Infeld-like
terms in the action.
In contrast to this, if one assumes all fields to be independent on
the extra
$x^5-$coordinate, solutions in the "pure" EYM model
are possible. These have been constructed in \cite{Volkov:2001tb} and
are spherically symmetric in four dimensions, extending
trivially into
one extra dimension. Generalisations of this model to $n$-extra
dimensions
have been constructed in \cite{Brihaye:2004kd}.

In this paper, we extend this model by introducing axial symmetry
in four dimensions.
Our five dimensional EYM solutions thus describe deformed vortex-type
solutions, which in the (3+1)-dimensional effective theory
can be interpreted as describing multimonopoles, respectively
monopole-antimonopole
pairs. Our Ansatz admits an interesting Kaluza-Klein picture in the
sense
that when boosting the solutions, we obtain new $d=4$ rotating and electrically
charged configurations.

Our paper is organized as follows: In Section 2, we give the model
including
the Ansatz and the boundary conditions. In Section 3, we describe our
numerical
results. In Section 4, we comment on the rotating solutions that we
obtain by boosting our solutions and in Section 5, we give our
conclusions.

\section{The model}
\subsection{Action principle  }
The five dimensional EYM-SU(2) system is described by the action
\begin{equation}
\label{action5}
I_5=\int d^{5}x\sqrt{-g_m }\Big(\frac{R }{16\pi G}
-\frac{1}{2g^2}Tr\{F_{MN }F^{MN} \}\Big),
\end{equation}
(throughout this letter, the indices $\{M,N,...\}$ (with $M$, $N$
running from one to five) will
denote the five dimensional
coordinates and $\{\mu,\nu,...\}$ the
coordinates of the four dimensional
physical spacetime;
the length of the extra dimension $x^5$ is taken to be one).

Here $G$ is the gravitational constant,
$R$ is the Ricci scalar associated with the
spacetime metric $g_{MN}$
and
$F_{MN}=\frac{1}{2} \tau^aF_{MN}^{(a)}$ is the gauge field strength
tensor defined as
$F_{MN} =
\partial_M A_N -\partial_N A_M + i[A_M , A_M  ],
$
where the gauge field is
$A_{M} = \frac{1}{2} \tau^a A_M^{(a)},$
$\tau^a$ being the Pauli matrices and $g$ the gauge coupling
constant.

Variation of the action (\ref{action5})
 with respect to  $g^{MN}$ and $A_M$ leads to the field equations
\begin{eqnarray}
\label{einstein-eqs}
R_{MN}-\frac{1}{2}g_{MN}R   &=& 8\pi G  T_{MN},
\\
\label{YM-eqs}
\nabla_{M}F^{MN}+i[A_{M},F^{MN}]&=&0,
\end{eqnarray}
where the YM stress-energy tensor is
\begin{eqnarray}
T_{MN} = 2{\rm Tr}
    ( F_{MP} F_{NQ} g^{PQ}
   -\frac{1}{4} g_{MN} F_{PQ} F^{PQ}).
\end{eqnarray}

\subsection{The ansatz}
In what follows we will consider vortex-type configurations,
assuming that both the matter functions and
the metric functions are
independent on the extra-coordinate $x^5$.
 Without any loss of generality, we consider a five-dimensional
metric parametrization
\begin{eqnarray}
\label{metrica}
ds^2 = e^{- a\psi }\gamma_{\mu \nu}dx^{\mu}dx^{\nu}
 + e^{ 2a\psi }(dx^5 + 2{\cal W}_{\mu}dx^{\mu})^2,
\end{eqnarray}
with $a=2/\sqrt{3}$.
With this assumption, the considered theory admits an interesting
Kaluza-Klein (KK) picture.
While the  KK reduction of the Einstein term in
(\ref{action5}) with respect to the Killing vector
$\partial/\partial x^5$
is standard,
for the  reduction of the YM action term,
it is convenient to take an  SU(2) ansatz
\begin{eqnarray}
\label{SU2}
A={\cal A}_{\mu}dx^{\mu}+g\Phi (dx^5+2 {\cal W}_\mu dx^\mu),
\end{eqnarray}
where ${\cal W}_\mu$ is a U(1) potential,
${\cal A}_{\mu}$ is a purely four-dimensional gauge field potential,
while  $\Phi$ corresponds after the dimensional reduction to a
triplet Higgs field.

This leads to the four dimensional action principle
\begin{eqnarray}
\label{action4}
I_4=\int d^{4}x\sqrt{-\gamma }\Big[
\frac{1}{4\pi G}\big(
\frac{\mathcal{R} }{4}
-\frac{1}{2}\nabla_{\mu}\psi \nabla^{\mu}\psi
-e^{2\sqrt{3}\psi}\frac{1}{4}G_{\mu \nu }G^{\mu \nu } \big)
-e^{2\psi/\sqrt{3}}\frac{1}{2g^2}Tr\{
{\cal F}_{\mu \nu }{\cal F}^{\mu \nu }\}
\\
\nonumber
-e^{-4\psi/\sqrt{3}}Tr\{ D_{\mu}\Phi D^{\mu}\Phi\}
- 2 e^{2\psi/\sqrt{3}}\frac{1}{g}G_{\mu \nu} Tr\{\Phi {\cal F}^{\mu
\nu} \}
-2e^{2\psi/\sqrt{3}} G_{\mu\nu}G^{\mu\nu}Tr\{  \Phi^2 \}
\Big],
\end{eqnarray}
where $\mathcal{R}$ is the Ricci scalar for the metric $\gamma_{\mu
\nu}$,
while
 ${\cal F}_{\mu \nu }=
\partial_{\mu}{\cal A}_{\nu}
-\partial_{\nu}{\cal A}_{\mu}+i [{\cal A}_{\mu},{\cal A}_{\nu}  ]$
and
 $G_{\mu \nu}=\partial_{\mu}{\cal W}_{\nu}-\partial_{\nu}{\cal
W}_{\mu}$
are the SU(2) and U(1) field strength tensors defined in $d=4$.

Here we consider five dimensional configurations possessing two more
Killing vectors
apart from $\partial/\partial x^5$,
$\xi_1=\partial/\partial \varphi$,
corresponding to an axially
symmetry of the four dimensional metric sector
(where the azimuth angle $\varphi$
 range from $0$ to $2 \pi$),
and $\xi_2=\partial/\partial t$,
with $t$ the time coordinate.

With these asumptions,
we consider the following parametrization of the four dimensional
line element
\begin{equation}
\label{metric}
d\sigma^2=\gamma_{\mu \nu}dx^{\mu}dx^{\nu}=\gamma_{tt}dt^2+d\ell^2=
- f(r,\theta)dt^2 +  \frac{m(r,\theta)}{f(r,\theta)}
(d r^2+ r^2 d \theta^2 )
           +  \frac{l(r,\theta)}{f(r,\theta)} r^2 \sin ^2 \theta
d\varphi^2,
\end{equation}
and the function $\psi(r,\theta)$ depending also on $r,\theta$ only.

The YM ansatz in this case
 is a straightforward generalization of the axially symmetric
 $d=4$ ansatz
obtained in the pioneering papers by Manton \cite{Manton:1977ht}
and Rebbi and Rossi \cite{Rebbi:1980yi},
and has been considered to some extend in \cite{Brihaye:2004kh}.
For the time and extra-direction
translational symmetry, we choose a gauge such that
$\partial A/\partial t=\partial A/\partial x^5=0$.
However, the
action of the Killing vector $\xi_1$
can be compensated by a gauge rotation
\begin{eqnarray}
\label{Psi}
{\mathcal{L}}_{\varphi} A_{N}=D_{N}\Psi,
\end{eqnarray}
with $\Psi$ being a Lie-algebra valued gauge function.
This introduces an winding number $n$ in the ansatz (which is a
constant of motion
and is restricted to be an integer)
and implies the existence of a potential $W$ with
\begin{eqnarray}
\label{relations}
F_{N \varphi} =& D_{N}W,
\end{eqnarray}
where $W=A_{\varphi}-\Psi$.

Thus, the most general axially symmetric 5D Yang-Mills ansatz
contains 15 functions: 12 magnetic
and 3 electric potentials and can be easily obtained in cylindrical
coordinates
$x^{i}=(\rho,\varphi,z$) (with $\rho=r\sin \theta,~z=r\cos\theta,$
and $r$, $\theta$ and $\varphi$ being
the usual spherical coordinates in (3+1)-dimensions)
\begin{eqnarray}
\label{A-gen-cil}
A_{N}=\frac{1}{2}A_{N}^{(\rho)}(\rho,z)\tau_{\rho}^n
        +\frac{1}{2}A_{N}^{(\varphi)}(\rho,z)\tau_{\varphi}^n
        +\frac{1}{2}A_{N}^{(z)}(\rho,z)\tau_{z},
\end{eqnarray}
where the only $\varphi$-dependent terms are the SU(2) matrices
(composed of the standard $(\tau_x,~\tau_y,~\tau_z)$ Pauli matrices)
$\tau_{\rho}^n=\cos n\varphi~\tau_x+\sin n\varphi~\tau_y,
~~
\tau_{\varphi}^n=-\sin n\varphi~\tau_x+\cos n\varphi~\tau_y.$
Transforming to
spherical coordinates, it is convenient to introduce,
without any loss of generality, a new SU(2) basis
$(\tau_{r}^n,\tau_{\theta}^n,\tau_{\varphi}^n)$,
with
$\tau_{r}^n=\sin \theta~\tau_{\rho}^n+\cos \theta~\tau_z,
~~
\tau_{\theta}^n=\cos \theta~\tau_{\rho}^n-\sin \theta~\tau_z,$
which yields
\begin{eqnarray}
\label{A-gen-sph}
A_{N}=\frac{1}{2}A_{N}^{(r)}(r,\theta)\tau_{r}^n
        +\frac{1}{2}A_{N}^{(\theta)}(r,\theta)\tau_{\theta}^n
        +\frac{1}{2}A_{N}^{(\varphi)}(r,\theta)\tau_{\varphi}^n.
\end{eqnarray}
For this parametrization
$2\Psi=n \tau_z =n \cos \theta  \tau_{r}^n
- n \sin \theta  \tau_{\theta}^n.$
The gauge invariant quantities expressed in terms of these functions
will be independent on the
angle $\varphi$.

Searching for solutions within the most general ansatz
is a difficult task.
Therefore we use in this paper a purely magnetic reduced ansatz with
six essential nonabelian potentials and
\begin{eqnarray}
\label{ansatz-spec}
\nonumber
A_{r}^{(r)}=A_{r}^{(\theta)}~=~A_{\theta}^{(r)}~=~A_{\theta}^{(\theta)}
~=~A_{\varphi}^{(\varphi)}=~A_{5}^{(\varphi)}=~A_{t}^{(a)}=0.
\end{eqnarray}
A suitable parametrization of the  nonzero
components of $A_N^{(a)}$ which factorizes the trivial
$\theta$-depencence
and admits a straightforward four dimensional picture is:
\begin{eqnarray}
\label{ansatz}
A_r^{(\varphi)}&=&\frac{1}{r}H_1(r,\theta),~~~A_{\theta}^{(\varphi)}=1-H_2(r,\theta),
~~~A_{\varphi}^{(r)}=-n \sin \theta H_3(r,\theta)+2g
J(r,\theta)\phi_1(r,\theta),
\\
\nonumber
A_{\varphi}^{(\theta)}&=&-n \sin \theta (1-H_4(r,\theta))+2g
J(r,\theta)\phi_2(r,\theta),
~~~A_{5}^{(r)}=\phi_1(r,\theta),~~~A_{5}^{(\theta)}=\phi_2(r,\theta),
\end{eqnarray}
(note that the $SO(3)$-symmetric ansatz is recovered for $H_1=H_3=\phi_2=J=0$ and
$H_2=H_4=\omega(r)$, $\phi_1=\phi(r)$).

To fix the residual abelian gauge invariance we choose the
gauge condition
\begin{eqnarray}
\label{gauge}
\nonumber
r \partial_r H_1 - \partial_\theta H_2 = 0.
\end{eqnarray}
We remark that
$A_{\varphi},~A_5$ have components along the same directions in
isospace.
Therefore, the $T_{\varphi}^5,~T_5^{\varphi}$ components of the
energy-momentum tensor
will be nonzero for axially symmetric YM configurations.
This implies the existence, in the five dimensional metric ansatz
(\ref{metrica}),
of one extradiagonal  $g_{5\mu}$ metric function, with
\begin{eqnarray}
{\cal W}_{\mu}=J(r,\theta)\delta_{\mu}^{\varphi}.
\end{eqnarray}

The $d=5$ EYM configurations  extremize also the
action principle (\ref{action4}) and can be viewed
as  solutions of the four dimensional
theory.
In this picture, $H_i(r,\theta)$ are the
 magnetic SU(2) gauge potentials, $\psi(r,\theta)$ is a dilaton,
$J(r,\theta)$
 is a U(1) magnetic potential,
 while $\phi_1(r,\theta),~\phi_2(r,\theta)$ are the components of a
Higgs field.
We mention also that, similar to the pure (E)-YMH case, we may define
a 't Hooft field strength tensor and an expression for the nonabelian
electric and magnetic charges
within the action principle (\ref{action4}).

\subsection{Boundary conditions}
\subsubsection{Metric functions}
To obtain  asymptotically flat regular solutions
with finite energy density
the metric functions have to satisfy the boundary conditions
\begin{equation} \label{b1}
\partial_r \psi|_{r=0}=\partial_r f|_{r=0}= \partial_r m|_{r=0}=
\partial_r l|_{r=0}= J|_{r=0}=0,
\end{equation}
which result from the requirement of regularity at the origin and
\begin{equation}\label{b2}
f|_{r=\infty}= m|_{r=\infty}=
l|_{r=\infty}=1,~\psi|_{r=\infty}=~J|_{r=\infty}=0,
\end{equation}
which result from the requirement of asymptotic flatness and finite
energy.
For solution with parity reflection symmetry (the case considered in
this paper),
the boundary conditions along the $z$ and $\rho$ axes are
(with $z=r \cos \theta$ and $\rho=r \sin \theta$)
\begin{equation}
\partial_\theta \psi|_{\theta=0,\pi/2}=
\partial_\theta J|_{\theta=0,\pi/2}=
\partial_\theta f|_{\theta=0,\pi/2} =
\partial_\theta m|_{\theta=0,\pi/2} =
\partial_\theta l|_{\theta=0,\pi/2} =0.
\end{equation}
Note that the boundary conditions for $f$, $m$, $l$, $\psi$
are similar to those derived in \cite{Brihaye:2002gp}, while the ones for $J$
are newly introduced.

\subsubsection{Matter functions}
A systematic study of the asymptotic behavior of the
$A_5$ component of the gauge field reveals that a  general enough
set of boundary conditions is given by
\begin{equation}
\label{phi-rinfty}
\lim_{r\to\infty}\phi_1=\eta \cos m\theta,~~~
\lim_{r\to\infty}\phi_2=\eta \sin m\theta\,,
\end{equation}
with $m=0,1,\dots$, and $\eta$ an arbitrary positive constant
($\eta=0$ implies $A_5=0$ which is outside the interest of this
paper).
This condition fixes the boundary conditions
at $r\to\infty$ for the other gauge potentials.
In deriving these conditions we use
the asymptotic analysis of the
Yang-Mills equations, requiring also the
finiteness of the total mass/energy, which
implies
that $F_{5 M}^{(a)}$ vanishes at infinity 
(see also \cite{Radu:2004ys} for a detailed discussion of 
this issue in a four dimensional EYMH theory).

 For even values of $m$ the asymptotic boundary conditions of the gauge functions $H_i$ are
\begin{eqnarray}
\label{evenm}
H_1=0,~~H_2=-m,~~
H_3=\frac{\cos \theta }{ \sin \theta}(\cos m\theta -1),~~
H_4=-\frac{\cos \theta}{\sin \theta}\sin m\theta\,.
\end{eqnarray}
while for odd $m$
\begin{eqnarray}
\label{oddm}
H_1=0,~~H_2=-m,~~
H_3=\frac{1}{\sin \theta}(\cos m \theta  -\cos \theta ),~~
H_4=-\frac{  \sin m\theta }{\sin \theta}.
\end{eqnarray}
In this paper we restrict ourselves to the simplest cases, $m=0$ and
$m=1$,
corresponding in a four dimensional picture to multimonopoles (MM)
and monopole-antimonopole (MA) configurations, respectively.

The boundary values at $r=0$ for $m=0$ are \cite{Hartmann:2001ic} :
\begin{eqnarray}
\label{r0MM}
H_{1}=H_3=0, ~~
H_{2}=H_{4}=1, ~~
\phi_{1}=\phi_{2}=0,
\end{eqnarray}
while for $m=1$, we impose \cite{Kleihaus:2000hx}:
\begin{eqnarray}
\label{r0MA}
H_{1}=H_3=0, ~~
H_{2}=H_{4}=1,~~
\cos \theta\, \partial_r\phi_{1}-\sin \theta\,\partial_r
\phi_{2}=0,~~
\sin \theta\, \phi_{1}+\cos \theta\, \phi_{2}=0,
\end{eqnarray}
which are the known conditions used in the study of 
four dimensional MM and MA  configurations, respectively.  
  The conditions along the axes are determined by the symmetries and 
finite energy density requirements. For $m=0$ solutions we impose  
\cite{Hartmann:2001ic}:
\begin{equation}
H_1|_{\theta=0,\pi/2}=H_3|_{\theta=0,\pi/2}=\phi_2|_{\theta=0,\pi/2}=0,~~
\partial_\theta H_2|_{\theta=0,\pi/2}
= \partial_\theta H_4|_{\theta=0,\pi/2}= \partial_\theta
\phi_1|_{\theta=0,\pi/2}
= 0,
\end{equation}
while the conditions satisfied by the $m=1$  configurations are  \cite{Kleihaus:2000hx}:
\begin{eqnarray}
H_1|_{\theta=0,\pi/2}=H_3|_{\theta=0,\pi/2}=
\partial_\theta H_2|_{\theta=0,\pi/2}=\partial_\theta
H_4|_{\theta=0,\pi/2}=0,
\\
\nonumber
\partial_\theta \phi_1|_{\theta=0}=\phi_1|_{\theta=\pi/2}=
\phi_2|_{\theta=0}=\partial_\theta \phi_2|_{\theta=\pi/2}=0.
\end{eqnarray}
In addition, regularity on the $z-$axis requires the conditions $l|_{\theta=0}=m|_{\theta=0}$,
 $H_2|_{\theta=0}=H_4|_{\theta=0}$ 
to be satisfied, for any values of the integers $(m,~n)$.

\subsection{Other relations}
 The assumed symmetries together with
the YM equations implies the following relations (we use here the relation (\ref{relations})
together with the Yang-Mills equations)
\begin{eqnarray}
\label{t1}
K&=&\int_V d^{3}x\sqrt{-g}~T_{\varphi}^5=2 Tr\{ \int_V
d^{3}x\sqrt{-g}
F_{M \varphi}F^{M t} \}
=2Tr\{\oint_{\infty}dS_{\mu}~\sqrt{-g}WF^{\mu 5} \},
\\
E_{h}&=&2 Tr\{\int_V d^3x  \sqrt{-g} F_{M 5}F^{M 5} \}=
2 Tr\{\oint_{\infty} d S_{\mu} \sqrt{-g} A_5 F^{\mu 5}  \},
\end{eqnarray}
where the volume integral is taken over the three dimensional
physical space.

These relations can easily be evaluated, by using the general sets of
boundary conditions
and the asymptotic expression
\begin{eqnarray}
\label{t2}
\phi_1\sim \eta\left(1-\frac{Q}{r}\right)\cos m\theta,
~~\phi_2\sim \eta\left(1-\frac{Q}{r}\right)\sin m\theta.
\end{eqnarray}
Thus we find
$E_h=4 \pi \eta^2 Q$, while $K=4 \pi n (1-(-1)^{m}) \eta Q$.

This implies that,
the magnitude of the gauge potentials $A_5$
should be nonzero at infinity, elsewhere $A_5\equiv 0$.
These relations  provide also an
useful tests to verify the accuracy
of the numerical calculation.

For the assumed asymptotic behavior, the mass of these solutions is
determined by the derivative of the metric function $f$
\begin{eqnarray}
\label{mass}
M=\frac{1}{2}\lim_{r \to \infty} r^2 \partial_r f.
\end{eqnarray}
When viewed as solutions of the four dimensional theory, the magnetic
charge of the
$m=0$ solutions is $n$, thus they corresponds to a generalization of
the gravitating
axially symmetric monopoles discussed in \cite{Hartmann:2001ic}.
In the same approach, the magnetic charge of the $m=1$ solutions is
zero (although locally the magnetic charge density is nonzero), thus
generalizing for a nonzero dilaton and U(1) field the known
monopole-antimonopole solutions
\cite{Kleihaus:2000hx}.

\newpage
\setlength{\unitlength}{1cm}

\begin{picture}(18,8)
\centering
\put(2,0.0){\epsfig{file=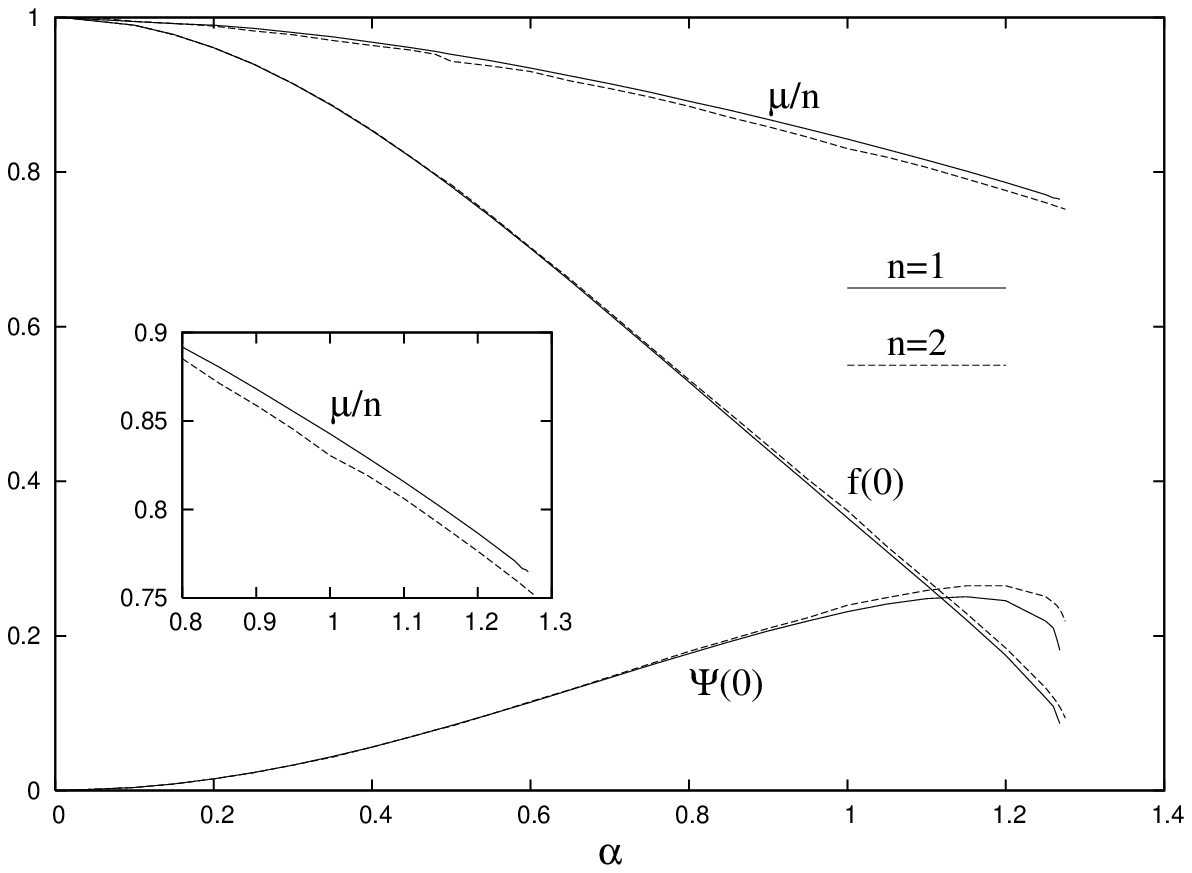,width=12cm}}
\end{picture}
\\
\\
{\small {\bf Figure 1.}
The dimensionless mass per the winding number $\mu/n$,
and the values of the metric functions $f(x)$ and $\psi(x)$
at the origin, $f(0)$, $\psi(0)$ are shown as functions of  $\alpha $
for $n=1,~2$ gravitating
solutions with $m=0$.}
\vspace{0.5cm}
\section{Numerical solutions}
By considering the rescalings $r \to r\eta g$ and $\phi \to
\phi/\eta$, the field
equations depend  only on the coupling
constants $\alpha=\sqrt{4\pi G}\eta$,
yielding the dimensionless mass   $\mu= (4\pi G\eta^2)^{-1}M$.

For $\alpha=0$ (no gravity) and no dependence on the $x^5$
coordinate, the four dimensional picture
 corresponds to the SU(2)-YMH theory in a fixed  Minkowski space.
Our solutions in this case describe $d=4$ nongravitating
multimonopoles (see e.g. \cite{Rebbi:1980yi})
and monopole-antimonopoles \cite{Kleihaus:1999sx}, respectively.
\subsection{$m=0$ "multimonopole" solutions}
We first constructed solution for $m=0$, $n \geq 1$
and varying $\alpha$.
Our numerical analysis
strongly suggests that the gravitating solutions
exist up to a ($n$-dependent) maximal value of $\alpha$,
$\alpha_{max}(n)$.
We find $\alpha_{max}(n=1)\approx 1.268$ and
$\alpha_{max}(n=2)\approx 1.275$. In the following, we will refer to
this
branch of solutions as the ``main branch''.

In Figure 1 some data characterizing the monopole solution
for $n=1$ and the multimonopole solutions for $n=2$ on this branch is
shown:
the mass per winding number $\mu/n$,
the value of the metric function $f(r)$ at the origin,
$f(0)$ and the value
of the dilaton field $\psi(r)$ at the origin, $\psi(0)$ are given as
functions
of $\alpha$.
As can be seen from this figure, the
mass ratio $\mu/n$ and the value $f(0)$  decrease for increasing
$\alpha$,
while
 $\psi(0)$ first increases
and then decreases starting from  $\alpha \sim 1.2$.

Note that $n=1$ in fact corresponds in this case to $SO(3)$-symmetric
solutions and thus implies $J(r,\theta)=0$. The solutions coincide
with the
ones obtained in \cite{Volkov:2001tb,Brihaye:2002gp}.
 Here, however we have
used isotropic coordinates as compared to Schwarzschild-like
coordinates used previously.
The numerical analysis in these latter papers has revealed
that several branches of solutions exist.
These branches (as illustrated of Figs. 1 and 3 of
\cite{Brihaye:2002gp}) have higher mass than the main branch.
It is very likely
(as suggested e.g. by the parameter $\psi(0)$ at the approach
of $\alpha_{max}$) that other branches of solutions appear
also for $n>1$. The construction of these branches turns out to be
numerically difficult and is not attempted in this publication.

In the case $n>1$, the function $J$ becomes non trivial although
it remain rather small, typically $\vert J \vert_{max}\sim 10^{-2}$
for $\alpha = \alpha_{max}$ in the case $n=2$.

\newpage
\setlength{\unitlength}{1cm}

\begin{picture}(18,7)
\centering
\put(2,0.0){\epsfig{file=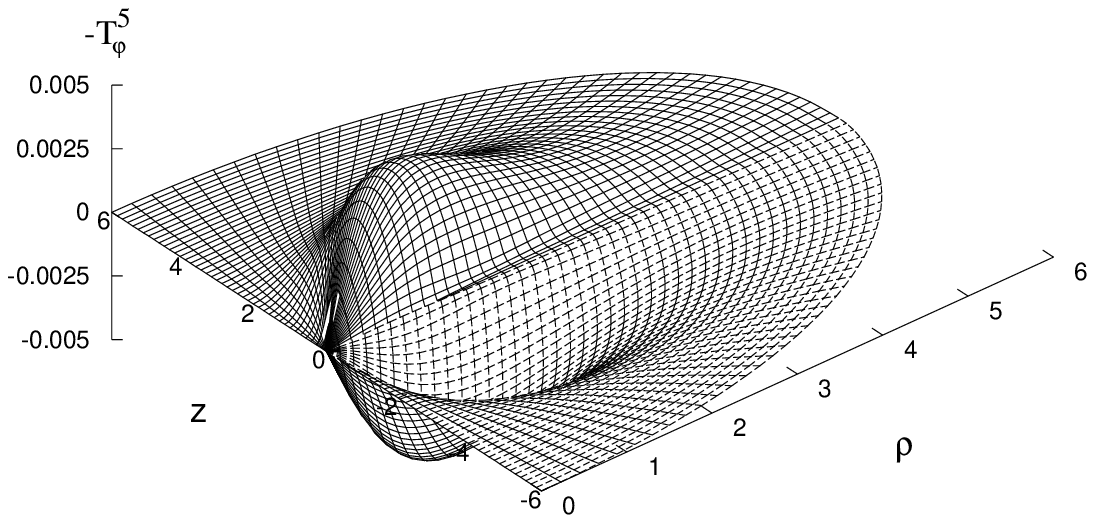,width=12cm}}
\end{picture}
\begin{picture}(19,5)
\centering
\put(2.6,0.0){\epsfig{file=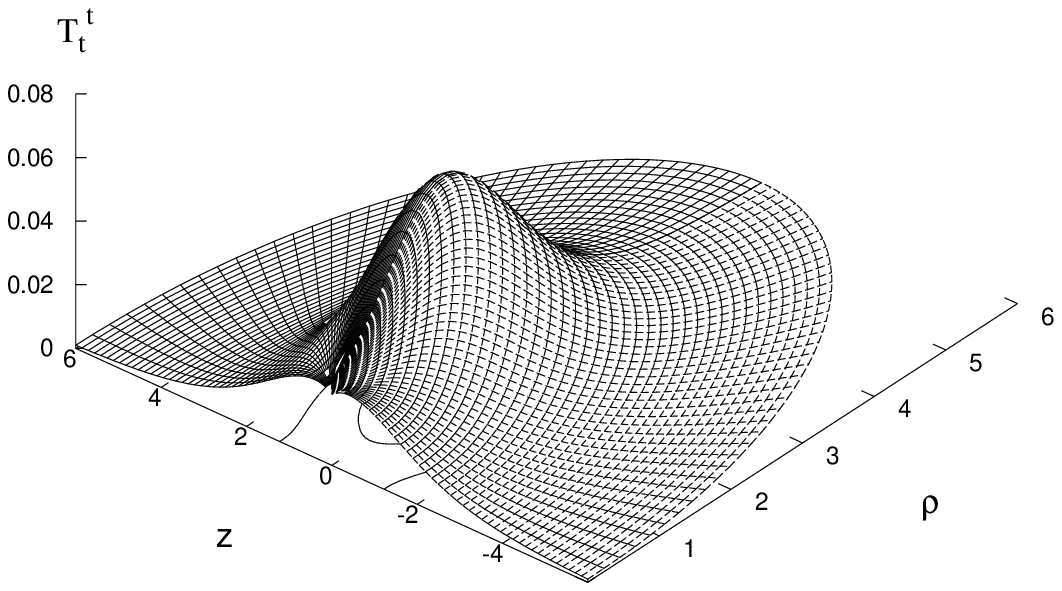,width=12cm}}
\end{picture}
\\
\\
{\small {\bf Figure 2.}
The components $T_{\varphi}^5$ and $T_t^t$ of the energy momentum
tensor are shown for
 a typical $m=0,~n=2$ solution, with $\alpha=0.95$.}
\vspace{0.5cm}
\\
Another feature of our solutions can be revealed
by studying the mass difference
$\Delta(n,\alpha) \equiv \mu_{n=1}- \mu_{n} /n$.
This quantity
characterizes the binding energy of the monopoles due to gravity.
It turns out that it is positive and we find typically,
for large $\alpha$ that  $\Delta(2,1.0) \sim 0.01$.
The binding energy values are very close (in fact of the same
order of magnitude within the numerical accurancy) to those
obtained in  a four-dimensional
EYMH-dilaton effective theory considered in
\cite{Brihaye:2002gp}, showing
that the supplementary function $J$ has a rather little influence
on the masses of the solutions.

In Figure 2 we show the energy density $\epsilon=-T_t^t$ and the
extradiagonal component $T^5_{\varphi}$ (which can be interpreted
as the momentum flux of the extra dimension across a surface
with $\varphi=const.$) of a typical $m=0$, $n=2$
solution
as function of the coordinates $z=r \cos\theta$ and $\rho=r
\sin\theta$
for a typical $n=2$ solution with $\alpha=0.95$.
As seen from this Figure,
the distributions of the mass-energy density
$-T_{t}^{t}$ can be different from those of spherical configurations,
showing a pronounced peak along the $\rho$-axis
and decreasing monotonically along the $z$-axis.
Equal density contours reveal a torus-like shape
of the solutions.
The picture is different for the $T^5_{\varphi}$-component which vanishes on the
 $\rho$-axis and changes the sign as $z \to -z$.

\subsection{$m=1$ "monopole-antimonopole" solutions}

A very different picture is found by taking $m=1$ in the asymptotic
boundary conditions (\ref{phi-rinfty}), (\ref{oddm})
(here we consider the case $n=1$ only).
When $\alpha$ is increased from zero,
a branch of $m=1$ solutions emerges from the uplifted version of the
$d=4$ flat spacetime MA
configurations. This branch ends at
a critical value $\alpha_{cr} \approx 0.65$.

\newpage
\setlength{\unitlength}{1cm}

\begin{picture}(18,10)
\centering
\put(2,0.0){\epsfig{file=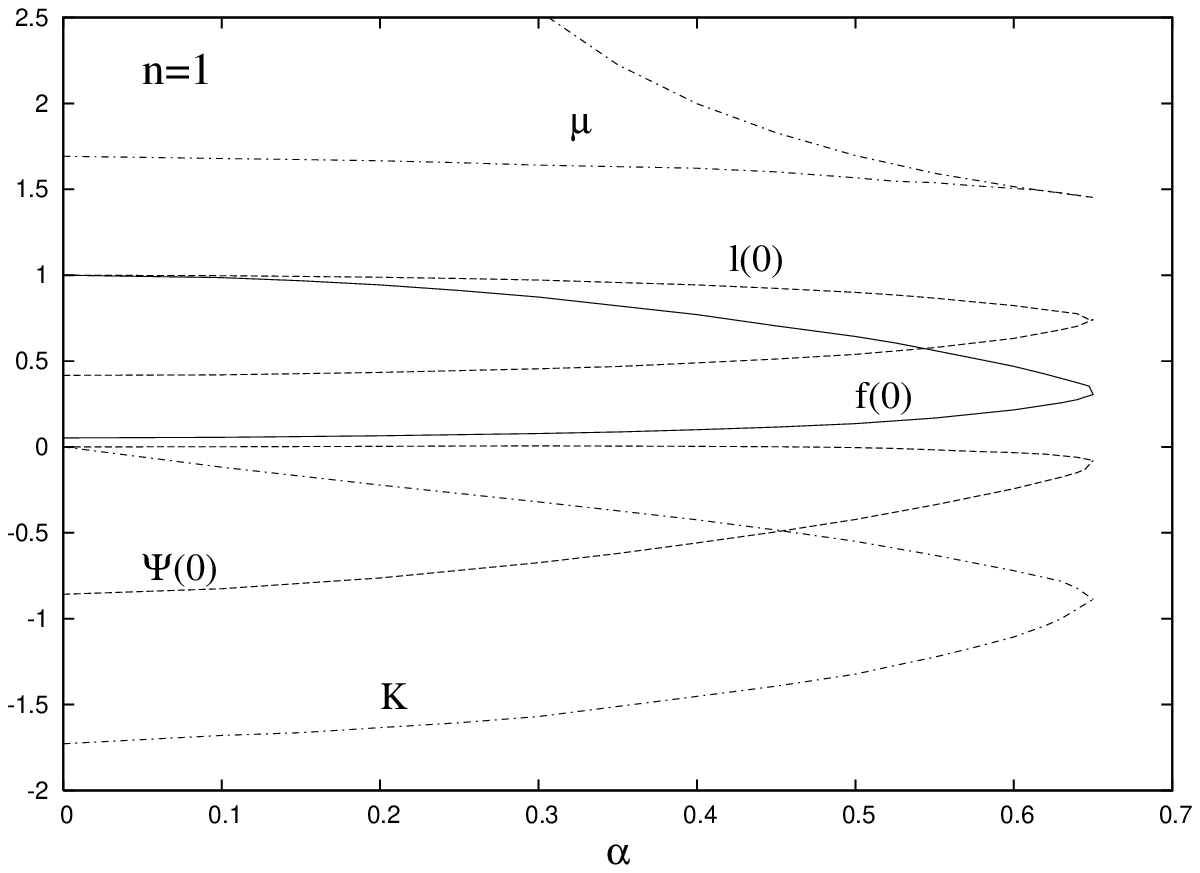,width=12cm}}
\end{picture}
\\
\\
{\small {\bf Figure 3.}
The dimensionless mass  $\mu$,
the value at the origin of the metric functions $f,~l$
and $\psi$ as well as the integral of the $T^5_{\varphi}$ component,
$K$,
are shown as functions of
$\alpha $ for $n=1$ gravitating solutions with $m=1$.}
\vspace{0.5cm}
\\
\\
As $\alpha \to \alpha_{cr}$, the geometry remains
regular with no event horizon appearing, and the mass  approaches a
finite value (see Figure 3).

Apart from this fundamental
branch, the $m=1$ solutions admit also
excited configurations, emerging in the $\alpha \to 0$ limit (after
a rescaling)  from the spherically symmetric solutions with $A_5=0$
(corresponding after dimensional reduction to solutions of a $d=4$
EYM-dilaton theory).
The
lowest excited branch, originating  from the one-node spherically
symmetric solution,
evolves smoothly from $\alpha=0$ to $\alpha_{cr}$ where it
bifurcates with
the fundamental branch.

Not surprisingly, the $m=1$ solutions share a number of
common properties with the  $d=4$ MA configurations in EYMH theory
discussed in
\cite{Kleihaus:2000hx}.
The functions $H_i, \phi_i$ and $f,l,m$
present a  shape similar to the  case considered in
\cite{Kleihaus:2000hx}. The energy density
$\epsilon = -T_{t}^t$ possesses maxima at $z=\pm d/2$ and a saddle
point at the origin, and presents the typical form exhibited in the
literature on MA solutions \cite{Kleihaus:2000hx, Kleihaus:1999sx,Radu:2004ys}.
The modulus of the fifth component of the gauge potential possesses
always two zeros at $\pm d/2$
on the $z-$symmetry axis.
 The excited solutions become infinitely
heavy as $\alpha \to 0$
while the distance $d$ tends to zero.
The   metric function $J(r,\theta)$ presents a nontrivial angular dependence, behaving
asymptotically as $J \sim J_0\sin^2 \theta/r$
(for $m=0$, $J$  decays as $1/r^2$ in the same limit).

The solutions mass, the integral (\ref{t1})
of the $T_{\varphi}^5$ component of the energy-momentum tensor
and the values of the metric functions $f$, $l$, $\psi$ at the origin
are plotted in Figure 3 as functions
of $\alpha$.
In Figure 4 we plot the  energy density $\epsilon=-T_t^t$ and the
extradiagonal component
$T_{\varphi}^5$
of a typical $m=1$ solution as a function of the
coordinates $\rho, z$, for $\alpha=0.45$.
Note the different shape of $T_{\varphi}^5$ as compared to the
$m=0$ case, which implies in this case a nonzero value of the volume integral (\ref{t1}).

Although we have restricted  the analysis  here to the simplest sets
of solutions, other excited $m=1$ branches
should exist
(these solutions have been found in the $d=4$ EYMH theory
\cite{Kleihaus:2000hx}).
These solutions do
not possess counterparts in flat spacetime and their  $\alpha
\to 0$ limit corresponds always to (higher nodes) vortex solutions of the
EYM  theory with $A_5=0$.

\newpage
\setlength{\unitlength}{1cm}

\begin{picture}(18,6)
\centering
\put(2,0.0){\epsfig{file=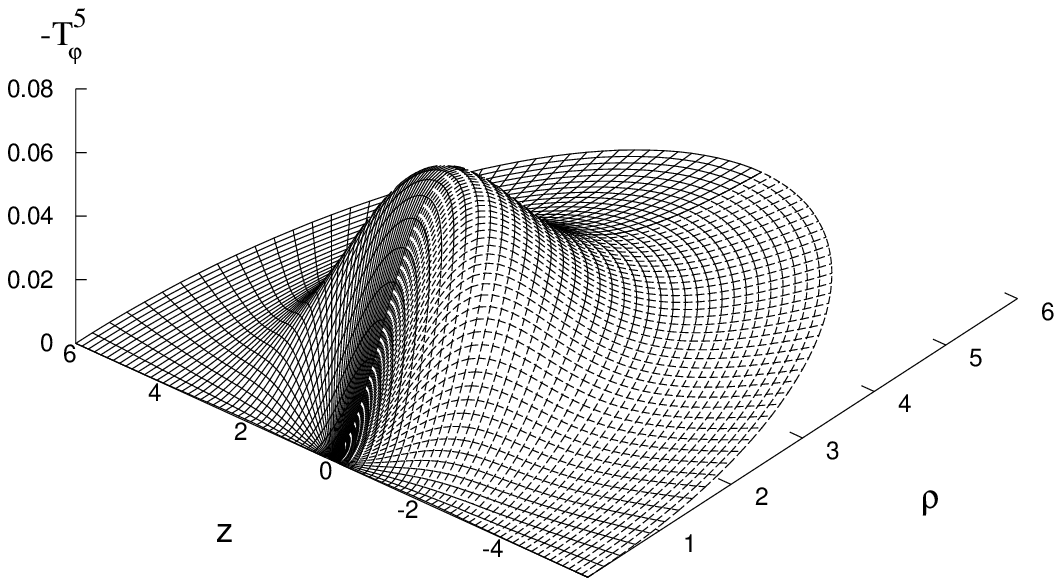,width=12cm}}
\end{picture}
\begin{picture}(19,5)
\centering
\put(2.6,0.0){\epsfig{file=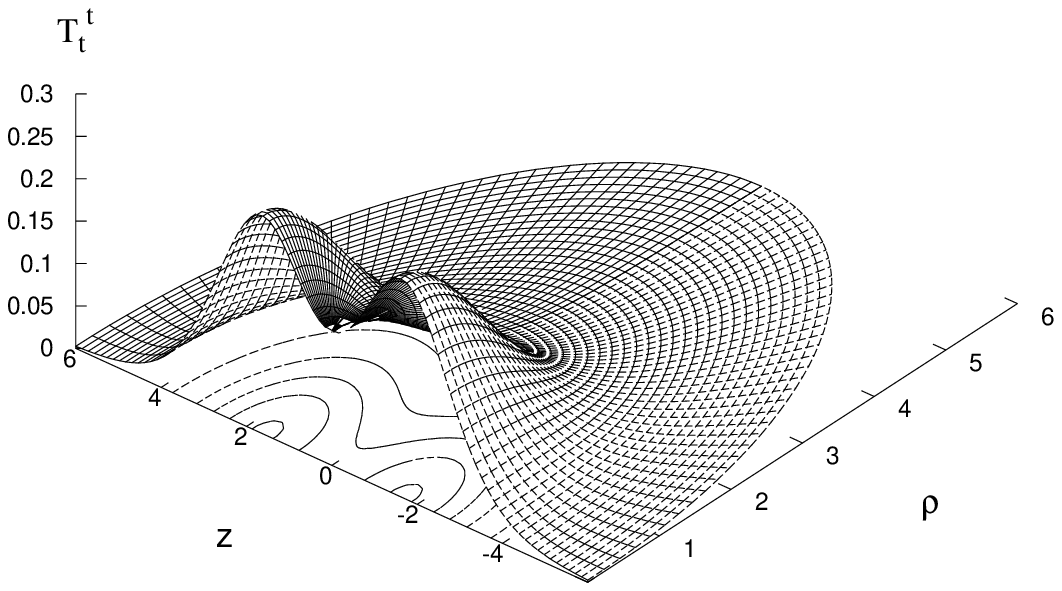,width=12cm}}
\end{picture}
\\
\\
{\small {\bf Figure 4.}
The components $T_{\varphi}^5$ and $T_t^t$ of the energy momentum
tensor are shown for
 a typical ($m=1,~n=1$) solution, with $\alpha=0.45$.}
\vspace{0.5cm}
%
\section{Remarks on rotating solutions}
As remarked in \cite{Hartmann:2000ja} there is a simple way to
generate electrically charged solutions
in a $d=4$ YMH theory without a Higgs potential, by starting
with a
pure magnetic configuration ($\vec A,\Phi_0$) and using the
transformation
$\vec A \to \vec A$, $ \Phi \to \Phi_0 \cosh \beta$,
$A_t \to \Phi_0 \sinh \beta$, with $\beta$ an arbitrary real
constant.
In a five dimensional picture, this corresponds to boosting in the
$(x^5,~t)$ plane a
purely magnetic YM vortex, according to
\begin{eqnarray}
\label{boost}
x^5=\cosh \beta~U+\sinh \beta ~T,~~t=\sinh \beta~U+\cosh \beta~T.
\end{eqnarray}
Also, for vacuum solutions extremizing ({\ref{action5}),
it has been known for some time that, by taking the product of the
$d=4$ Schwarzschild
solution with a circle and
boosting it in the fifth direction, the entire family of electrically
charged
(magnetically neutral) KK  black holes is generated.

A more complicated picture is found in the presence of nonabelian
matter fields.
However, it can be proven that,
given a $d=5$ initial configuration $(\psi,~{\cal
W}_{\varphi},~\gamma_{\mu \nu},~A_N)$,
with two Killing vectors $\partial/\partial t,~\partial/\partial
x^5$,
by applying

the coordinate transformation (\ref{boost}),
the new form of the EYM solution is given by the line element
\begin{eqnarray}
\label{metrica-new}
ds^2 = e^{- a\bar{\psi} }\bar{\gamma}_{\mu \nu}dx^{\mu}dx^{\nu}
 + e^{ 2a\bar{\psi} }(dU + 2\bar{{\cal W}}_{\varphi}d\varphi
 +2\bar{{\cal W}}_{T}dT)^2,
\end{eqnarray}
the same $SU(2)$ potentials $A_r,~A_\theta,~A_\varphi$ and
\begin{eqnarray}
\label{tr1}
A_T=\sinh \beta(\phi_1 \frac{\tau_r^n}{2}+\phi_2
\frac{\tau_r^\theta}{2}),~~
A_U=\cosh \beta(\phi_1 \frac{\tau_r^n}{2}+\phi_2
\frac{\tau_r^\theta}{2}).
\end{eqnarray}
The new quantities in (\ref{metrica-new}) are defined by
\begin{eqnarray}
\label{cant-new}
\nonumber
e^{2a \bar{\psi}}&=&e^{2a\psi}\cosh^2
\beta+e^{-a\psi}\gamma_{tt}\sinh^2 \beta,
\\
\nonumber
\bar{{\cal W}}_{\varphi}&=&\frac{e^{2a\psi}\cosh \beta~{\cal
W}_{\varphi}}
{e^{2a\psi}\cosh^2 \beta+e^{-a\psi}\gamma_{tt}\sinh^2 \beta},
\\
\bar{{\cal
W}}_{T}&=&\frac{1}{2}\frac{(e^{2a\psi}+e^{-a\psi}\gamma_{tt})\sinh \beta \cosh \beta}
{e^{2a\psi}\cosh^2 \beta+e^{-a\psi}\gamma_{tt}\sinh^2 \beta},
\\
\nonumber
\bar{\gamma}_{\mu \nu}dx^{\mu}dx^{\nu}&=&
e^{a(\bar{\psi}-\psi)}\left(\gamma_{tt}(dT-2\sinh \beta {\cal
W}_{\varphi}
d \varphi)^2+d \ell^2 \right).
\end{eqnarray}
For the metric ansatz (\ref{metric}), we find
\begin{eqnarray}
\label{tr2}
\bar{f}=\frac{f e^{a\psi}}{\sqrt{e^{2a\psi}\cosh^2
\beta-e^{-a\psi}f\sinh^2 \beta}},~~\bar{m}=m,~~~~\bar{l}=l.
\end{eqnarray}
The following "reality condition" follows straightforward
\begin{eqnarray}
\label{tr3}
e^{2a\psi}\cosh^2 \beta-e^{-a\psi}f\sinh^2 \beta>0
\end{eqnarray}
which turn out to be satisfied by all considered configurations
(although we could not find an analytical argument).

The dimensional reduction of these configurations along the
$U$-direction
provides new solutions in the $d=4$ EYMH-U(1)-dilaton theory.
As different from the original configurations,
the boosted configurations
 present a nonzero $\gamma_{\varphi T}$ term, thus corresponding
to rotating electrically charged solutions. The angular momentum
density of these $d=4$ configurations is given by $\cosh
\beta~T_{\varphi}^5$ and has the typical shape presented in Figures
2, 4. However, although they will rotate locally, the total angular
momentum of the MM solutions is zero, and the spacetime consists in two
regions rotating in opposite directions  \footnote{ In
Einstein-Maxwell theory, a zero total angular momentum implies a
static configuration. The situation may be different for a more
general matter content. The existence of rotating solution of
Einstein equation with a vanishing ADM angular momentum has been
noticed in \cite{Herrera}.}. The situation is different for MA
configurations, whose ADM angular momentum becomes proportional with
the constant $K$ of the static solutions as defined by (\ref{t1}).

\section{Conclusions}
In this paper we consider axially symmetric vortex-type solutions in
the
$d=5$ EYM-SU(2) theory. Our motivation is two-fold: firstly, such
solutions are
interesting in their own right.
Secondly, it has been shown that, after dimensional reduction, the
system
corresponds to a particular EYMH-U(1)-dilaton model.
Thus we may hope to learn more about physics of the gravitating YMH
model in four dimensions.
This is interesting especially in connection with the question of
gravitating rotating solutions.
In this context,
we have presented a simple procedure to generate $d=4$ rotating solutions
with nonabelian matter fields
starting with static $d=5$ EYM vortex-type solutions.

The $d=4$ rotating solutions we find by taking $m=1$ in the general
set of boundary conditions generalize the  MA configurations
discussed recently in  \cite{Paturyan:2004ps} and may help to
clarify the issue of the limiting solutions, left unsolved in that
paper. Even more interesting is the $m=0$ case. As yet to the best
of our knowledge there is no example of globally regular rotating
nonabelian solutions with a nonvanishing magnetic charge presented
in the literature. However, the boosted $m=0$ solutions correspond
in $d=4$ to globally regular dyons solutions with a nonzero
extradiagonal metric component $\gamma_{
\varphi T}$ associated with rotation. 
Although these configurations rotate locally, their
global angular momentum vanishes as predicted in
\cite{Brihaye:2004kh, vanderBij:2002sq}. This  suggests that similar
configurations should exist also in EYMH theory.

 We expect also that $d=5$ EYM  theory possesses a
whole sequence of  solutions, obtained within the ansatz
(\ref{ansatz}) for an arbitrary $m >1$.
By boosting these solutions, new   $d=4$ configurations
describing rotating chains and vortex rings \cite{Kleihaus:2004fh} 
can be generated.

We close by pointig out another possible interpretation of the solutions discussed in this paper.
Since $A_t=0$ in the matter ansatz (\ref{ansatz-spec}) and there is also
 no time dependence, our configurations will solve also the $d=5$ EYM equations on the
Euclidean section, obtained by analytically continuing $t \to i
\tau$ (with an arbitrary periodicity of $\tau$ for these regular
solutions).
Now, the KK reduction with respect to the Killing vector $\partial/\partial \tau$
corresponds in a four dimensional picture to rotating regular instantons in a
 EYM-dilaton theory. The U(1) and Higgs field are zero in this case while
$\phi_i$ corresponds to electric SU(2) potentials.

More details on these globally regular solutions
will be given elsewhere.
\\
\\
{\bf\large Acknowledgements} \\
YB is grateful to the
Belgian FNRS for financial support.
The work of ER is carried out
in the framework of Enterprise--Ireland Basic Science Research
Project
SC/2003/390 of Enterprise-Ireland.
%
%
%
%
%


\end{document}